\documentclass[final,5p,times,twocolumn]{elsarticle}

\usepackage{graphicx}
\usepackage{amssymb}
\journal{Journal of Alloys and Compounds}
\begin{document}
\begin{frontmatter}



\title{Cell dynamics modeling of phase transformation and metastable phase formation}
\author{Masao Iwamatsu}
\ead{iwamatsu@ph.ns.tcu.ac.jp}

\address{Department of Physics, School of Liberal Arts, Tokyo City University, Setagaya-ku, Tokyo 158-8557, Japan}

\begin{abstract}
The phase transition kinetics in three phase systems was investigated using the numerically efficient cell dynamics method.  A phase-field model with a simple analytical free energy and single order parameter was used to study the kinetics and the thermodynamics of a three-phase system.  This free energy is able to achieve three phases coexistence, which for simplicity we call $\alpha$, $\beta$ and $\gamma$ phases.  Our study focused on the kinetics of phase transition rather than the nucleation of a seed of a new phase that was introduced into the matrix of the old phase when the relative stability of the three phases were changed.  We found dynamical as well as kinetically arrested static scenarios in the appearance of the macroscopic metastable phase.  A few other interesting scenarios of the kinetics of phase transition in this three-phase system will be demonstrated and discussed.
\end{abstract}

\begin{keyword}
Cell dynamics \ Phase transformation \ Metastable phase


\end{keyword}
\end{frontmatter}


\section{\label{sec:sec1}Introduction}
\label{sec:level1}
The study of the phase transition in a multi-phase system has been studied for many years. In particular, the multi-step phase transition which involves the formation of intermediate long-lived metastable third phase when the phase transition from one phase to another occurs has been studied for more than a century~\cite{Ostwald1897,Cahn1969}.  Recently, a renewed interest in the phase transition in the multi-phase system~\cite{Poon2002,Vekilov2004,Sears2007,Toth2007} has emerged.  The formation of the thermodynamically metastable third phase is important academically as well as industrially because many industrial products are in a long-lived metastable state~\cite{Poon2002}.

In this report, we use the cell dynamics method employed previously to study the kinetics of the phase transition in a simple three phase system~\cite{Iwamatsu2005b} and study the kinetics for more wide spectrum of free energy landscape using phase-field model, which has been extensively used to study various scenarios of phase transition~\cite{Valls1990,Steinbach1996,Raabe1998,Pusztai2008,Emmerich2009}.  We employ cell dynamics method as we want to consider the evolution of multiple nucleus with general symmetry while the previous authors, instead, considered the evolution of one-dimensional traveling wave solution.~\cite{Bechhoefer1991,Celestini1994,Granasy2000,Evans1997a,Evans1997b} of single nucleus.  We pay special attention to the formation of metastable third phase during the phase transition between first to second phases.

\section{Model free energy for three-phase system}
\label{sec:sec2}

In order to study the phase transition kinetics in the multi-phase system, we will use the partial differential equation called isothermal phase-field equation for the non-conserved order parameter~\cite{Castro2003}:
\begin{equation}
\frac{\partial \psi}{\partial t}=-\frac{\delta \mathcal{F}}{\delta \psi}
\label{eq:2-1}
\end{equation}
where $\psi$ is called phase-field and is actually the {\it non-conserved} order parameter and $\mathcal{F}$ is the free energy functional (grand potential), which is usually written as the square-gradient form:
\begin{equation}
\mathcal{F}[\psi]=\int \left[\frac{1}{2}(\nabla \psi)^{2}+f(\psi)\right]d{\bf r} 
\label{eq:2-2}
\end{equation}
where the local part of the free energy $f(\psi)$ will be specified later and will realize the multi-phase system.  As a special solution, Eq.(\ref{eq:2-1}) is expected to allow an interface-controlled growth with nearly constant velocity~\cite{Chan1977}.  It should be noted that these dynamics Eq.~(\ref{eq:2-1}) always guarantee that the total free energy decreases monotonically because~\cite{Langer1992}
\begin{equation}
\frac{d\mathcal{F}}{dt}
=\int \frac{\delta \mathcal{F}}{\delta \psi}\frac{\partial \psi}{\partial t}dr
=-\int\left(\frac{\delta \mathcal{F}}{\delta \psi}\right)^{2}dr\leq 0
\label{eq:2-2x}
\end{equation}
by reducing the local free energy $f$ or the surface tension that is proportional to $(\nabla \psi)^{2}$ in Eq.~(\ref{eq:2-2}).

\begin{figure}[htbp]
%Fig.1
\begin{center}
\includegraphics[width=0.6\linewidth]{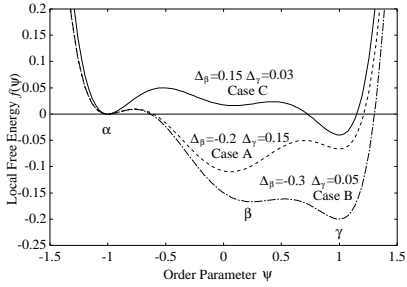}
\caption{
The model triple-well free energy $f(\psi)$.  In this figure, we show three cases: Case A ($\Delta_{\beta}=-0.3$, $\Delta_{\gamma}=0.05$)$f_{\alpha}<f_{\beta}<f_{\gamma}$, Case B ($\Delta_{\beta}=-0.2$, $\Delta_{\gamma}=0.15$) $f_{\alpha}<f_{\beta}<f_{\gamma}$, and Case C ($\Delta_{\beta}=0.2$, $\Delta_{\gamma}=0.05$) $f_{\alpha}<f_{\beta}<f_{\gamma}$.}
\label{fig:1}
\end{center}
\end{figure}

We will consider the simplest multi-phase system with only three phases and introduce a local free energy function $f(\psi)$ of triple-well form as a function of the non-conserved order parameter $\psi$.  We extend the analytical free energy $f(\psi)$ proposed by Widom~\cite{Widom1978} in order to achieve the relative stability of three phases:
\begin{equation}
f(\psi) = \frac{1}{4}(\psi+1)^{2}(\psi-1)^{2}(\psi^{2}+\Delta_{\beta})+\Delta_{\gamma}\left(\frac{1}{3}\psi^{3}-\psi\right)-\frac{2}{3}\delta
\label{eq:1-1}
\end{equation}
which can realizes the two-phase and three-phase system according to the magnitude of two parameters $\Delta_{\beta}$ and $\Delta_{\gamma}$.  We have three phases $\alpha$ with the free energy $f_{\alpha}$ at $\psi_{\alpha}=-1$, $\beta$ with $f_{\beta}$ at $\psi_{\beta}\sim 0$, and $\gamma$ with $f_{\gamma}$ at $\psi_{\gamma}=1$.  Now the free energy landscape consists of two wells $\alpha$ and $\gamma$-wells, or three wells $\alpha$, $\beta$ and $\gamma$-wells.  The parameter $\Delta_{\beta}$ controls the relative stability of the intermediate $\beta$ phase while the parameter $\Delta_{\gamma}$ controls the stability of the $\gamma$ phase.  Because the role of $\gamma$ and $\alpha$ is interchangeable, we will only consider the cases when the $\gamma$ phase or the intermediate $\beta$ phase are the most stable thermodynamic phase.  Therefore, we will restricted to $\Delta_{\gamma}>0$ as the free energy at the $\gamma$ phase is given by
\begin{equation}
f_{\gamma} = f(\psi=1)=-\frac{4}{3}\Delta_{\gamma}
\label{eq:1-2}
\end{equation}
while the free energy of the $\alpha$ phase is always fixed to $f_{\alpha}=0$.  The free energy of $\beta$ phase is approximately given by
\begin{equation}
f_{\beta} \sim f(\psi=0)=\frac{1}{4}\Delta_{\beta}-\frac{2}{3}\Delta_{\gamma}.
\label{eq:1-2x}
\end{equation}
Several typical shapes of the free energy function $f(\psi)$ for several sets of the parameters $\Delta_{\beta}$ and $\Delta_{\gamma}$ are shown in Fig.~\ref{fig:1}.  

\begin{figure}[htbp]
\begin{center}
\includegraphics[width=0.6\linewidth]{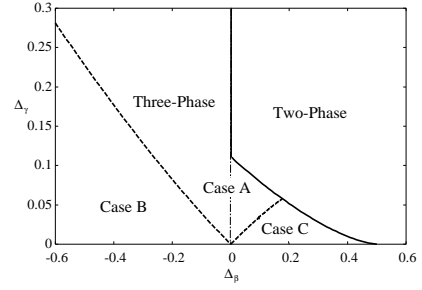}
\caption{
The phase diagram for the model free energy Eq.~(\ref{eq:1-1}).  Case A:$f_{\beta}<f_{\gamma}<f_{\alpha}$, Case B:$f_{\gamma}<f_{\beta}<f_{\alpha}$, and Case C:$f_{\gamma}<f_{\alpha}<f_{\beta}$. The triple point is at $\Delta_{\beta}=0$ and $\Delta_{\gamma}=0$, where all three phases coexist with $f_{\alpha}=f_{\beta}=f_{\gamma}=0$.}
\label{fig:2}
\end{center}
\end{figure}

We observe from the figure~\ref{fig:1} that there are typically three cases for the free energy landscape according to the relative stability of three states.  Case A when $f_{\beta}<f_{\gamma}<f_{\alpha}$ is a rather special case as the intermediate $\beta$ phase is most stable.  Case B when $f_{\gamma}<f_{\beta}<f_{\alpha}$ has been considered by Bechhoefer et al.~\cite{Bechhoefer1991} and Celestini and ten Bosch~\cite{Celestini1994} though they considered only a special traveling solution.  Case C when $f_{\gamma}<f_{\alpha}<f_{\beta}$ is the most interesting and relevant to the phase transition of several soft-condensed matter system~\cite{Poon2002,Vekilov2004} and has been studied by Evans et al.~\cite{Evans1997a,Evans1997b} for the kinetics of conserved order parameter and by the present author~\cite{Iwamatsu2005b} for the non-conserved order parameter when $f_{\alpha}=f_{\gamma}$.  The phase diagram of this system in the two parameter space $(\mu,\delta)$ is shown in Fig.~\ref{fig:2}.  

Similar triple-well potentials of different functional forms were used by several workers to study the nucleation~\cite{Granasy2000,Bradley1989} and the metastable phase formation using phase-field model of non-conserve~\cite{Iwamatsu2005b,Bechhoefer1991,Celestini1994,Granasy2000} and conserved~\cite{Evans1997a,Evans1997b} order parameter.
\section{Results of numerical simulation using cell dynamics}
\label{sec:sec3}
\subsection{Cell dynamics and phase-field model}

According to the cell dynamics method~\cite{Oono1988}, the partial differential equation (\ref{eq:2-1}) is transformed into the finite difference equation in space and time:
\begin{equation}
\psi(t+1,n)=F[\psi(t,n)]
\label{eq:2-3}
\end{equation}
where the time $t$ is discrete integer and the space is also discrete and is expressed by the site index (integer) $n$.  The mapping $F$ is given by
\begin{equation}
F[\psi(t,n)]=g(\psi(t,n))+\frac{1}{2}\left[<<\psi(t,n)>>-\psi(t,n)\right]
\label{eq:2-4}
\end{equation}
where the definition of $<<*>>$ for the two-dimensional square grid is given by
\begin{equation}
<<\psi(t,n)>>=\frac{1}{6}\sum_{i=\mbox{nn}}\psi(t,i)
+\frac{1}{12}\sum_{i=\mbox{nnn}}\psi(t,i)
\label{eq:2-5}
\end{equation}
with "nn" means the nearest neighbors and "nnn" the next-nearest neighbors of the square grid. Instead of the original map function $g\left(\psi\right)=\psi-1.3\tanh\psi$~\cite{Oono1988}, we used the map function that is directly derived from the free energy landscape $f(\psi)$:
\begin{equation}
g\left(\psi\right)=\psi-\frac{df}{d\psi}
\label{eq:2-6}
\end{equation}
which is essential in order to study the kinetics of phase transition when a subtle balance of the relative stability of the three phases in the three-phase system plays a crucial role~\cite{Iwamatsu2005b,Ren2001a,Iwamatsu2005a}.

\subsection{Kinetics of phase transition in a three-phase system}
Since we are most interested in the evolution or regression of the metastable phase during the phase transformation after nucleation, we will only examine the kinetics of phase transition when various composite nuclei made from three phases are prepared in the materials using the cell dynamics equation Eq.~(\ref{eq:2-3}) and the model free energy defined by Eq.~(\ref{eq:1-1}) (Fig.~\ref{fig:1}).

\subsubsection{Case A ($f_{\beta}<f_{\gamma}<f_{\alpha}$)}
We have incorporated the above free energy Eq.~(\ref{eq:1-1}) into the cell-dynamics code Eq.~(\ref{eq:2-3}) used previously to study the KJMA (Kolmogorov-Johnson-Mehl-Avrami) dynamics~\cite{Iwamatsu2008a}.  In this case, the intermediate middle $\beta$-well of the free energy landscape in Fig.~\ref{fig:1} is most deep and the left and the right well is shallower. In this case the $\beta$ phase is most stable and the $\gamma$ phase and the $\alpha$ phase are metastable.

\begin{figure}[htbp]
\begin{center}
\includegraphics[width=0.6\linewidth]{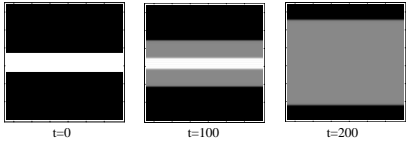}
\caption{
A Gray-level view of the time evolution of a three-layer stripes when $\Delta_{\beta}=-0.2$ and $\Delta_{\gamma}=0.15$ (Case A).  The black area is the metastable $\alpha$ phase, the gray area is the stable $\beta$ phase, and the white area is the metastable $\gamma$ phase.  Initially the less stable $\gamma$ phase is sandwiched by metastable $\alpha$ phase at $t=0$.  Even though, the initial state does not contain the stable $\beta$ phase, it appears spontaneously at the metastable $\alpha$-$\gamma$ interface and continues to grow.}
\label{fig:3}
\end{center}
\end{figure}

In figure~\ref{fig:3}, we prepare the metastable (white) $\gamma$ strip of width 21 embedded in the middle of the metastable (black) $\alpha$ phase with size $128\times 128$ and observe the evolution of the system.  Throughout this paper, we will use $128\times 128$ system where a periodic boundary condition is imposed.  We use the potential parameters $\Delta_{\beta}=-0.2$ and $\Delta_{\gamma}=0.15$ which correspond to the curve for Case A in Fig.~\ref{fig:1}.  Figure ~\ref{fig:3} shows that the most stable (gray) $\beta$ phase appears spontaneously at the interface of two metastable $\gamma$ and $\alpha$ phases during the evolution.  This stable $\beta$ slab continues to grow and invades the metastable $\gamma$ as well as $\alpha$ phase.  Finally the whole material transforms into the stable $\beta$ phase.

\subsubsection{Case B ($f_{\gamma}<f_{\beta}<f_{\alpha}$)}
Now the right $\gamma$ well is deepest, and the middle $\beta$ well is the next and the left $\alpha$ well is most shallow (Fig.~\ref{fig:1}).  This staircase configuration of the free energy landscape was previously studied by Bechhoefer et al.~\cite{Bechhoefer1991} and Celestini and ten Bosch~\cite{Celestini1994}.

\begin{figure}[htbp]
\begin{center}
\includegraphics[width=0.6\linewidth]{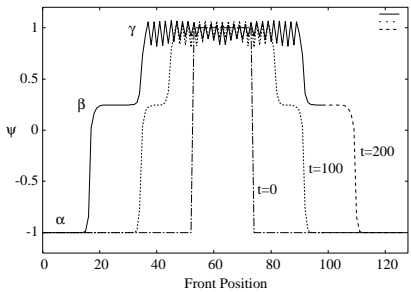}
\caption{
The cross section view of the stripe structure similar to Fig.~\ref{fig:3}.  The most stable $\gamma$ phase with $\psi=1$ is sandwiched by the least stable $\alpha$ phase with $\psi=-1$.  In this case, dynamically stable $\beta$ phase with $\psi\simeq 0$ appears even though it is thermodynamically metastable.}
\label{fig:4}
\end{center}
\end{figure}

\begin{figure}[htbp]
\begin{center}
\includegraphics[width=0.6\linewidth]{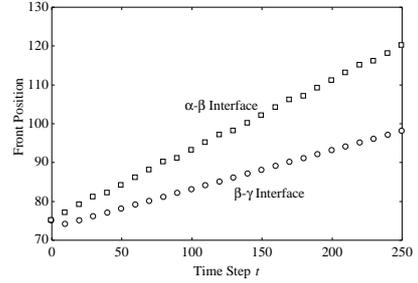}
\caption{
The time evolution of the position of $\alpha$-$\beta$ interface and $\beta$-$\gamma$ interface.  The position is measured from the bottom of the $128\times 128$ pixels of Fig.~\ref{fig:4}.  Two nearly straight lines indicate the propagation with constant velocities that are given by the slopes.}
\label{fig:5}
\end{center}
\end{figure}

In Fig. ~\ref{fig:4} we start from the same stripe configuration as that in Fig.~\ref{fig:3}.  Now the moving $\alpha$-$\gamma$ interface unbind into a pair of moving $\alpha$-$\beta$ and moving $\beta$-$\gamma$ interface and a macroscopic slab of metastable $\beta$ phase appears.  The small oscillation in the phase field $\psi$ in the $\gamma$ phase is probably due to the interference effect because it is confined.  Since the $\beta$-$\gamma$ interfacial velocity is slower than the $\alpha$-$\beta$ velocity (Fig.~\ref{fig:5}), the metastable $\beta$ slab appears and continues to grow.

Figure \ref{fig:5} shows the time evolution of the position of the $\alpha$-$\beta$ interface defined by $\psi=-0.5$ and the $\beta$-$\gamma$ interface defined by $\psi=0.5$.  Two nearly straight lines with different slopes indicate that the two interfaces propagate with different constant velocities.  Since the $\beta$-$\gamma$ free energy difference is smaller than the $\alpha$-$\beta$ free energy difference (Fig.~\ref{fig:1}), the $\beta$-$\gamma$ interfacial velocity is slower than the $\alpha$-$\beta$ interfacial velocity (Fig.~\ref{fig:5}) as they are expected to be proportional to the free energy difference~\cite{Chan1977,Iwamatsu2005a}.

\begin{figure}[htbp]
\begin{center}
\includegraphics[width=0.6\linewidth]{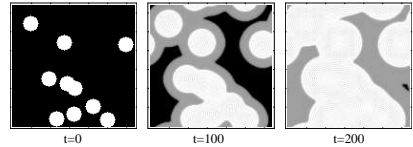}
\caption{
Evolution of the most stable circular nuclei of $\gamma$ phase (white) embedded in least stable $\alpha$ phase (black).  Again the dynamically stabilized macroscopic metastable $\beta$ layer (gray) appears around the $\gamma$ core and grows at the $\gamma$-$\alpha$ interface.}
\label{fig:6}
\end{center}
\end{figure}

Our numerical simulation directly confirmed the existence of a dynamically stabilized metastable state predicted by the theoretical calculations of Bechhoefer et al.~\cite{Bechhoefer1991} and Celestini and ten Bosch~\cite{Celestini1994}.  However, the appearance of this dynamically stabilized metastable $\beta$ phase is not due to the special symmetry of planer wave front assumed by those authors~\cite{Bechhoefer1991,Celestini1994}.  Figure \ref{fig:6} show the evolution of the multiple of circular nucleus of most stable $\gamma$ phase embedded in least stable $\alpha$ phase.  Again, the metastable $\beta$ layer appears around the growing $\gamma$ core spontaneously and will grow.

\subsubsection{Case C ($f_{\gamma}<f_{\alpha}<f_{\beta}$)}
In this case, the right $\gamma$-well is deepest, while the left $\alpha$-well is the next, and the middle $\beta$ well is the shallowest and least stable (Fig~\ref{fig:1}).  This free energy configuration with metastable intermediate phase or buried metastable phase was previously studied by a few authors~\cite{Cahn1969,Iwamatsu2005b,Evans1997a,Evans1997b}.

\begin{figure}[htbp]
\begin{center}
\includegraphics[width=0.45\linewidth]{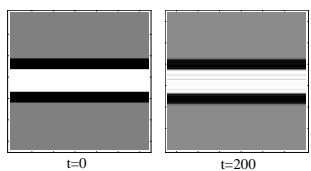}
\caption{
A composite nucleus of stable $\gamma$ stripe (white) sandwiched by a less stable $\alpha$ phase (black) embedded in most metastable $\beta$ phase (gray) at $t=200$.  This configurations is kinetically arrested and cannot grow even though there is a seed of the most stable $\gamma$ phase (white).}
\label{fig:7}
\end{center}
\end{figure}

Now the direct phase transition from the metastable right $\alpha$ phase to the stable left $\gamma$ phase (Fig.~\ref{fig:1}) is prohibited from Eq.~(\ref{eq:2-2x}) as the reaction path must go through the intermediate $\beta$ phase with higher free energy.  As a consequence, a composite nucleus of a stripe made from the most stable $\gamma$ phase of width 21 sandwiched by the next stable $\alpha$ phase of width 33 embedded in the center of least stable $\beta$ phase as shown in Fig.~\ref{fig:7} will be kinetically arrested as the most stable $\gamma$ phase cannot invade the $\alpha$ phase and cannot grow.  Even though the $\alpha-\gamma$ front is freeze, the $\alpha-\beta$ front can grow as the $\alpha$ phase can grow by consuming the surrounding $\beta$ phase.  However, since we do not include thermal fluctuation (noise) in our kinetic equation (\ref{eq:2-2}), the $\alpha-\beta$ front is virtually arrested as there exists a small barrier between $\alpha$ well and $\beta$ well (Fig.~\ref{fig:1}).

\begin{figure}[htbp]
\begin{center}
\includegraphics[width=0.6\linewidth]{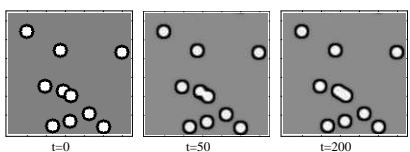}
\caption{
Composite nuclei with the most stable circular core of $\gamma$ phase surrounded by a thin layer of less stable $\alpha$ phase embedded in least stable $\beta$ phase.  Only a fusion of two colliding nucleus is observed near the center.  This boiled-egg configuration is stable and kinetically arrested.}
\label{fig:8}
\end{center}
\end{figure}

The appearance of this kinetically arrested metastable $\beta$ phase around the composite nucleus made from the most stable $\gamma$ core wrapped by the least stable $\alpha$ skirt is not due to the planer symmetry.  Figure \ref{fig:8} shows the evolution of the multiple of circular composite nuclei consist of most stable $\gamma$ phase of radius 5 surrounded by a thin layer of least stable $\alpha$ phase of radius 8 embedded in a least stable $\beta$ phase.  Again, these composite nuclei are kinetically arrested and cannot grow.  Only a fusion of two colliding nuclei occurs near the center.  The resulting oval nucleus of $\gamma$ wrapped by a thin layer of metastable $\alpha$ is static and stable again.  Then the metastable $\beta$ phase environment survives even though there are seeds of most stable $\gamma$ phase.  This configuration called {\it boiled-egg} structure was predicted from experiments~\cite{Poon2002,Poon1999a,Renth2001} and was found by the cell-dynamics simulation of three-phase system by the author~\cite{Iwamatsu2005b} for the special case of single composite nucleus and the $\alpha$-$\gamma$ equilibrium.

Since we start from the static configuration, not only the evolution of the most stable $\gamma$ phase by $\alpha\rightarrow \gamma$ transition is prohibited but also the evolution of the surrounding $\alpha$ phase by $\beta\rightarrow \alpha$ transition cannot be initiated spontaneously as the free energy barrier between $\alpha$ phase and $\beta$ phase exists (Fig.~\ref{fig:1}).  Naturally, the introduction of the thermal noise initiates $\beta\rightarrow\alpha$ transition~\cite{Iwamatsu2005b}.  However, this kinetically arrested phase implies fairly long lifetime of these metastable configuration~\cite{Iwamatsu2005b}.

\section{Conclusion}
\label{sec:sec4}
In this paper, we have used the cell dynamics method to explore various scenario of phase transition in three-phase system that is characterized by a single non-conserved order parameter.  We have considered three cases when a relative stability of three phases, $\alpha$, $\beta$ and $\gamma$ phases are different.  We have found several interesting scenarios of the kinetic of phase transition in a three-phase system: (a) the appearance of stable third phase from the interface of two metastable phase, (b) the appearance of dynamically stable third metastable phase from the moving interface between stable and metastable phase, and (c) the appearance of kinetically arrested third metastable phase around the composite nuclei made from the stable phase surrounded by a thin layer of metastable phase.  Although, the last two examples have already been predicted theoretically~\cite{Bechhoefer1991,Celestini1994}, or numerically~\cite{Iwamatsu2005b} for special configuration, we showed in this paper using cell-dynamics simulation that these predictions are qualitative correct even for multiple nuclei in two-dimensions and for more wide variety of free energy landscape.


\begin{thebibliography}{99} %% The number "99" means that this list has more than nine items.
\bibitem{Ostwald1897} W. Ostwald, Z. Phys. Chem. (Munich) 22 (1897) 286.
\bibitem{Cahn1969} J.W. Cahn, J. Am. Ceram. Soc. 52 (1969) 118.
\bibitem{Poon2002} W.C.K. Poon, J. Phys.: Condens. Matter 14 (2002) R859.
\bibitem{Vekilov2004} P.G. Vekilov, Cryst. Growth. Des. 4 (2004) 671.
\bibitem{Sears2007} R.P. Sears, J. Phys.: Condens. Matter 19 (2007) 033101.
\bibitem{Toth2007} G.I. T\'oth and L. Gr\'an\'asy, J. Chem. Phys. 127 (2007) 074709, 074710. 
\bibitem{Iwamatsu2005b} M. Iwamatsu, Phys. Rev. E 71 (2005) 061604.
\bibitem{Valls1990} O.T. Valls and G.F. Mazenko, Phys. Rev. B 42 (1990) 6614.
\bibitem{Steinbach1996} I. Steinbach, F. Pezzola, B. Nestler, M. Seesselberg, R. Prieler, G.J. Scmitz, and J.L.L. Rezende, Physica D 94 (1996) 135.
\bibitem{Raabe1998} D. Raabe, {\it Computational Material Sciece} (Wiley-VCH, Weinheim, 1998), chap. 10.
\bibitem{Pusztai2008} T. Pusztai, G. Tegze, G.I. T\'oth, L. {K\"ornyei}, G. Bansel, Z. Fan, and L. Gr\'an\'asy, J. Phys.: Condens. Matter 20 (2008) 404205.
\bibitem{Emmerich2009} H. Emmerlich, J. Phys.: Condens. Matter 21 (2009) 464103.
\bibitem{Bechhoefer1991} J. Bechhoefer, H. {L\"owen}, and L.S. Tuckerman, Phys. Rev. Lett. 67 (1991) 1266.
\bibitem{Celestini1994} F. Celestini and A. ten Bosch, Phys. Rev. E 50 (1994) 1836.
\bibitem{Granasy2000} L. Gr\'an\'asy and D.W. Oxtoby, J. Chem. Phys. 112 (2000) 2410.
\bibitem{Evans1997a} R.M.L. Evans, W.C.K. Poon, and M.E. Cates, Europhys. Lett. 38 (1997) 595.
\bibitem{Evans1997b} R.M.L. Evans and M.E. Cates, Phys. Rev. E 56 (1997) 5738.
\bibitem{Castro2003} M. Castro, Phys. Rev. B 67 (2003) 035412.
\bibitem{Chan1977} S-K. Chan, J. Chem. Phys. 67 (1977) 5755.
\bibitem{Langer1992} J.S. Langer, in {\it Solids Far From Equilibrium}, edited by C. Godr\`eche (Cambridge UP, Cambride, 1992), chap. 3.
\bibitem{Widom1978} B. Widom, J. Chem. Phys. 68 (1978) 3878.
\bibitem{Bradley1989} R.M. Bradley and P.N. Strenski, Phys. Rev. B 40 (1989) 8967.
\bibitem{Oono1988} Y. Oono and S. Puri, Phys. Rev. A 38 (1988) 434; S. Puri and Y. Oono, {\it ibid} 1542.
\bibitem{Ren2001a} S.R. Ren, I.W. Hamley, P.I.C. Teixeira, and P.D. Olmsted, Phys. Rev. E 63 (2001) 041503.
\bibitem{Iwamatsu2005a} M. Iwamatsu and M. Nakamura, Jpn. J. Appl. Phys. Part 1 44 (2005) 6688.
\bibitem{Iwamatsu2008a} M. Iwamatsu, J. Chem. Phys. 128 (2008) 084504.
\bibitem{Poon1999a} W.C.K. Poon, F. Renth, R.M.L.  Evans, D.J. Fiarhurst, M.E. Cates, and P.N. Pusey, Phys. Rev. Lett. 83 (1999) 1239.
\bibitem{Renth2001} F. Renth, W.C.K. Poon, and R.M.L. Evans, Phys. Rev. E 64 (2001) 031402.
\end{thebibliography}
\end{document}